\documentclass[aps,twocolumn]{revtex4}
\usepackage{graphicx}

\begin{document}
\newcommand{\be}{\begin{equation}}
\newcommand{\en}{\end{equation}}
\title{Corrections to the Entropy in Higher Order Gravity}
\author{Luis Alejandro Correa-Borbonet}\email{borbonet@cpd.ufmt.br}
\affiliation{Departamento de F\'\i sica \\ Universidade Federal de Mato Grosso,\\
Av. Fernando Corr\^{e}a da Costa, $s/n^o$-Bairro Coxip\'{o}\\
78060-900-Cuiab\'{a}-MT, Brazil}

\begin{abstract}
Thermal corrections to the entropy of black holes in the Lovelock gravity are calculated.  As the  thermodynamic behavior of the black holes of this theory falls into two classes, the thermodynamic quantities are computed in each case. Finally, the logarithmic prefactors are obtained in two different limits.

\end{abstract}

\maketitle
\section{Introduction}
During the late $1960$'s and early $1970$'s an intense research activity in the field 
of black hole physics lead to the discovery of the laws of black hole mechanics. Soon, it was
realized a striking resemblance with the laws of thermodynamics\cite{beke,hawking}. 
In this context Bekenstein proposed 
that the entropy in black holes is proportional to the surface area of the event horizon. Moreover, Hawking  found that black holes generate also a thermal radiation, due to quantum pair production in their gravitational potential gradient and the presence of the event horizon. The temperature of this radiation is
given in terms of the surface gravity at the event horizon, $T_{H}=\hbar \kappa/2\pi$. An immediate consequence of this identification of the temperature is that the proportio\-nality constant between the entropy and the area is fixed, $S_{BH}=A/4G$.

Althought this knowledge about the black hole thermodynamics is 
well established, we do not have a complete answer to this question in the statistical mechanics framework.
However, there has been a substantial progress in identifying the microscopic degrees of freedom responsible for the Bekenstein-Hawking entropy. This advance has come from string theory\cite{vafa} and loop quantum gravity\cite{ashtekar}. In the case of string theory it is present a massless spectrum including the graviton, and at low energy it gives supergravity effec\-tives theories. Black Holes therefore appear as classical solutions of low energy string theory.

The next step in this line of research has been the study of the leading correction to entropy. A common characteristic of the different approaches is the proportionality to $\mathrm{ln}\, S_{BH}$
\cite{kaul},\cite{carlip},\cite{solodukin}. However, the proportionality cons\-tant does not exhibit the same universality. It can be shown that logarithmic corrections to thermodynamic entropy arise in all thermodynamic systems when small stable fluctuations around equilibrium are taken into account\cite{das}. The stability condition is equivalent to the specific heat being 
positive. 

On the other hand the study of the thermodynamic properties of black holes has been 
extended to higher order gravity theories\cite{myers}. 
Within these theories there is a special class of gravitational actions, of 
higher order in the curvature, known as Lovelock gravity
\cite{lovelock}. Lovelock gravity is exceptional in the sense that
although it contains higher powers of the curvature in the Lagrange
density, the resulting equations of motion contain no more 
than second derivatives of the metric. It is also a covariant and ghost free 
theory as it happens in the case of Einstein's General Relativity. 

An important result that was found in the thermodynamic context is that
the area  law is a peculiarity of the Einstein-Hilbert theory \cite{scan}. 
These facts motivates a deeper study of the thermodynamics of 
the black hole solutions of such exotic theories \cite{setare}\cite{odintsov},\cite{eaacb}. 
In this paper we will study the 
corrections to the entropy for the black holes solutions of Lovelock gravity.
We shall first briefly review such a formulation. 

\section{Higher Dimensional Gravity}
The Lanczos-Lovelock action is a polynomial of degree $[d/2]$ in the curvature,
which can be expressed in the language of forms as \cite{scan}
\be
I_{G}=\kappa \int \sum_{m=0}^{[d/2]}\alpha _{m}L^{(m)},  \label{eq:accion}
\en
where $\alpha _{m}$ are arbitrary constants, and $L^{(m)}$ is given by
\be
L^{(m)}=\epsilon _{a_{1}\cdots a_{d}}R^{a_{1}a_{2}}\!\cdot \!\cdot \!\cdot
\!R^{a_{2m-1}a_{2m}}e^{a_{2m+1}}\!\cdot \!\cdot \!\cdot \!e^{a_{d}},
\label{lp}
\en
with $R^{ab}$ being the Riemann curvature two-forms given by
\be
R^{ab}=d\omega^{ab}+\omega^{a}_{c}w^{cb}\quad .
\en
Here $w_{ab}$ are the spin connection one-forms and $e^{a}$ the vielbein. 
A wedge product between forms is understood throughout.

The corresponding field equations can be obtained varying with respect to
$e^{a}$ and $w^{ab}$. In \cite{scan} the expression for the coefficients
$\alpha_{m}$ was found requiring the existence of a unique cosmological
constant. In such a case these theories are described by the action 
\be
I_{k}=\kappa \int \sum_{p=0}^{k}c_{p}^{k}L^{(p)}\;,  \label{eq:Itk}
\en
which corresponds to (\ref{eq:accion})  with the choice
\be
\alpha _{p}:=c_{p}^{k}=\left\{
\begin{array}{ll}
\frac{l^{2(p-k)}}{(d-2p)}\left(
\begin{array}{c}
k \\
p
\end{array}
\right)  & ,
\, p\leq k \\
0 & ,
\, p>k
\end{array}
\right.   \label{Coefs}
\en
for the parameters, where  $1 \leq k \leq [(d-1)/2]$.
For a given dimension $d$, the coefficients $c^{k}_{m}$ give rise to a
family of inequivalent theories, labeled by $k$ which represent the
highest power of curvature in the Lagrangian. This set of theories
possesses only two fundamental constants,  $\kappa$ and $l$, related
respectively to the gravitational constant $G_{k}$ and the cosmological
constant $\Lambda$ through 
\be
\kappa =\frac{1}{2(d-2)\Omega_{d-2}G_{k}}\quad ,
\en
\be
\Lambda=-\frac{(d-1)(d-2)}{2l^{2}}\quad .
\en
For black hole solutions that are
asymptotically flat we consider the vanishing cosmological constant limit
case. When $l\rightarrow \infty$ the only non-vanishing terms in
Eq(\ref{eq:Itk}) is the 
kth one; therefore the action is obtained from Eq(\ref{eq:accion}) with the 
choice of coefficients 
\be
\alpha _{p}:=\tilde{c}_{p}^{k}=\frac{1}{(d-2k)}\delta _{p}^{k}\; ,
\label{Coefs0}
\en
in this case the  action reads
\be
\tilde{I}_{k}\!=\!\frac{\kappa }{(d-2k)}\!\!\int \!\!\epsilon _{a_{1}\cdots
a_{d}}R^{a_{1}a_{2}}\!\cdot \!\cdot \!\cdot
\!R^{a_{2k-1}a_{2k}}e^{a_{2k+1}}\!\cdot \!\cdot \!\cdot \!e^{a_{d}}\quad .
\label{ActionSL}
\en
Note that for $k=1$ the Einstein action without cosmological constant is
recovered, while for $k=2$ we obtained the Gauss-Bonnet action,
\be
I_{2}=\frac{(d-2)!\kappa}{(d-4)}\int d^{d}x \sqrt{-g}(-R_{\mu \nu \alpha 
\beta}R^{\mu \nu \alpha \beta}+4R_{\mu \nu}R^{\mu
\nu}-R^{2})\quad . \label{eq:lagran} 
\en
Returning to the action (\ref{eq:Itk}) we remember that this set of theories possess asymptotically 
AdS black hole solutions given by\cite{scan}, 
\begin{eqnarray}
ds^{2} & = & -\left(1+\frac{r^{2}}{l^{2}}-\left(\frac{2G_{k}M+\delta_{d-2k,1}}{r^{d-2k-1}}\right)^{1/k}\right)dt^{2}+\nonumber\\
 & &\frac{dr^{2}}{1+\frac{r^{2}}{l^{2}}-\left(\frac{2G_{k}M+
 \delta_{d-2k,1}}{r^{d-2k-1}}\right)^{1/k}}+r^{2}d\Omega^{2}_{d-2}\quad . \label{eq:bhl}
\end{eqnarray}
The black hole mass for any value of $k$ is a monotonically increasing function of the horizon radius 
$r_{+}$, which reads
\be
M(r_{+})=\frac{r^{d-2k-1}_{+}}{2G_{k}}\left(1+\frac{r^{2}_{+}}{l^{2}}\right)^{k}-\frac{1}{2G_{k}}\delta_{d-2k,1}.
\label{eq:masa}
\en
The presence of the Kronecker delta within the metrics $(\ref{eq:bhl})$ signals the exis\-tence of two possible black hole vacua($M=0$) with different causal structures. The generic case, with $d-2k\neq 1$, 
is 
\begin{eqnarray}
ds^{2} & = & -\left(1+\frac{r^{2}}{l^{2}}-\left(\frac{2G_{k}M}{r^{d-2k-1}}\right)^{1/k}\right)dt^{2}+\nonumber\\
 & &\frac{dr^{2}}{1+\frac{r^{2}}{l^{2}}-\left(\frac{2G_{k}M
 }{r^{d-2k-1}}\right)^{1/k}}+r^{2}d\Omega^{2}_{d-2}\quad . \label{eq:sads}
\end{eqnarray}
Analogously with the Schwarzschild-AdS metric, this set possesses a conti\-nuous mass spectrum, whose vacuum state is the AdS spacetime. The other case is obtained for odd dimensions, and it is a peculiarity of Chern-Simons theories. From 
$(\ref{eq:bhl})$ we obtain, 
\begin{eqnarray}
ds^{2} & = & -\left(1+\frac{r^{2}}{l^{2}}-\left(2G_{k}M+1\right)^{1/k}\right)dt^{2}+\nonumber\\
 & &\frac{dr^{2}}{1+\frac{r^{2}}{l^{2}}-\left(2G_{k}M+1\right)^{1/k}}+r^{2}d\Omega^{2}_{d-2}.
 \label{eq:csbh}
\end{eqnarray}
Here, the black hole vacuum differs from AdS spacetime.
\section{Canonical Formalism}
In this section we review\cite{das} the derivation of the entropy in the canonical formalism. Looking for the entropy corrections it is considered the existence of small thermal fluctuations around the equilibrium. Then, we begin with the canonical partition function  
\be
Z(\beta)=\int^{\infty}_{0}\Omega(E)e^{-\beta E}dE,
\en
where $\Omega(E)$ is the density of states, that can be obtained from the partition function doing an inverse
Laplace transform
\be
\Omega(E)=\frac{1}{2\pi i}\int^{c+i\infty}_{c-i\infty}Z(\beta)e^{\beta E}d\beta=\frac{1}{2\pi i}\int^{c+i\infty}_{c-i\infty} e^{S(\beta)}d\beta, \label{eq:density}
\en
where
\be
S(\beta)=\mathrm{ln} Z(\beta)+\beta E
\en
is the entropy. The integral can be performed by the method of steepest descent around the saddle point $\beta_{0}=1/T_{0}$ such that $S'_{0}=(\partial S/\partial \beta)_{\beta=\beta_{0}}=0$. Here $T_{0}$ is the
equilibrium temperature. Expanding the entropy around $\beta_{0}$, we have
\be
S=S_{0}+\frac{1}{2}(\beta-\beta_{0})^{2}S''_{0}+...\label{eq:exp}
\en 
Substituting (\ref{eq:exp}) in (\ref{eq:density}) and integrating we obtain 
\be
\Omega(E)=\frac{e^{S_{0}}}{\sqrt{2\pi S''_{0}}}\quad .
\en
Finally, using the Boltzmann's formula, is obtained 
\be
\mathcal{S}=\mathrm{ln}\, \Omega=S_{0}-\frac{1}{2}\mathrm{ln}\left(S''_{0}\right)\label{eq:correc}+.....
\en
Here $\mathcal{S}$ is the entropy at equilibrium. This is to be distinguished from the function $S(\beta)$, which is the entropy at any temperature.
 
The logarithmic term can be transformed taking into account that $S''_{0}$ is the fluctuation of the 
mean squared energy, i.e,
\be
S''_{0}=<E^{2}>-<E>^{2},
\en
and that the specific heat is $C=(\partial E/\partial T)_{T_{0}}$. Therefore
\be
\mathcal{S}=S_{0}-\frac{1}{2}\mathrm{ln}\left(C  T^{2}\right).\label{eq:correc}
\en
This result apply to stable thermodynamic systems with small fluctuations around the equilibrium. The stability condition is equivalent to the specific heat being positive. On the other hand is assumed that the quantum 
fluctuations of the thermodynamics quantities under consideration are small. In other words, for black holes very 
close to extremality ($T\rightarrow 0$), the fluctuation analysis ceases to be valid due to large quantum fluctuactions\cite{das}.
\section{Corrections to the Entropy in Higher Order Gravity}
\subsection{Asymptotically AdS black hole solutions}
Reviewing the thermodynamic properties of the black hole solution (\ref{eq:sads}) we began 
with the expression for the Hawking temperature, that is
\be
T_{H}=\frac{1}{4\pi k_{B}k}\left((d-1)\frac{r_{+}}{l^{2}}+\frac{d-2k-1}{r_{+}}\right),\label{eq:tempt}
\en
where $r_{+}$ is the horizon radius. Note that for all $k$ such that $d-2k-1\neq 0$ the temperature has the same behavior that the Schwarzschild-AdS black hole(k=1), that is: the temperature diverges at $r_{+}=0$. Also has a minimum at $r_{c}$ given by
\be
r_{c}=l\sqrt{\frac{d-2k-1}{d-1}},
\en
and grows linearly for large $r_{+}$.
Consequently we can calculate the specific heat $C_{k}=\frac{\partial M}{\partial T}$ as a function of $r_{+}$. Using (\ref{eq:tempt}) and (\ref{eq:masa}) we obtain,
\be
C_{k}=k\frac{2\pi k_{B}}{G_{k}}r^{d-2k}_{+}\left( \frac{r^{2}_{+}+r^{2}_{c}}{r^{2}_{+}-r^{2}_{c}}\right)\left(1+\frac{r^{2}_{+}}{l^{2}}\right)^{k-1}.
\en
Here the function $C_{k}$ has an unbounded discontinuity at $r_{+}=r_{c}$, signaling a phase transition. We will deal with black hole with horizon radius that satisfies the condition $r_{+}>r_{c}$, where the specific heat is positive and the correction formula (\ref{eq:correc}) can be apply. 

Finally we present the entropy function
\be
S_{k}=k\frac{2\pi k_{B}}{G_{k}}\int^{r_{+}}_{0}r^{d-2k-1}\left(1+\frac{r^{2}}{l^{2}}\right)^{k-1}dr,
\en
obtained from the Euclidean path integral formalism\cite{scan}. Similar results are obtained in the 
Lagrangian formalism \cite{olea}.

For simplicity we just perform the calculations for black holes with $k=2$. Therefore, for the entropy 
we get
\be
S^{(0)}_{2}=\frac{4\pi k_{B}}{G_{2}}r^{d-4}_{+}\left[\frac{1}{(d-4)}+\frac{r^{2}_{+}}{(d-2)l^{2}} \right]
\en
where $(0)$ stands for the uncorrected entropy.
In terms of this entropy the Hawking temperature and the specific heat are given by: 
\begin{eqnarray}
T_{H}& = &\left[\frac{1}{8\pi k_{B}l^{2}}\left(\frac{4\pi k_{B}}{G_{2}l^{2}}\right)^{-1/(d-2)} \right] \times \\ \nonumber
& &\times \frac{\left[(d-1)+\frac{(d-5)l^{2}}{r^{2}_{+}}\right]}{\left[\frac{l^{2}}{(d-4)r^{2}_{+}}
+\frac{1}{d-2}\right]^{1/(d-2)}}
\quad (S^{(0)}_{2})^{1/(d-2)},
\end{eqnarray}
\be
C_{2}=\left(\frac{r^{2}_{+}+r^{2}_{c}}{r^{2}_{+}-r^{2}_{c}}\right)\left(1+\frac{l^{2}}{r^{2}_{+}}\right)
\frac{1}{\frac{l^{2}}{(d-4)r^{2}_{+}}+\frac{1}{(d-2)}}S^{(0)}_{2}.
\en
In the limit $\mathrm{r}_{+}\gg l$, $C_{2}$ approaches the value
\be
C_{2}=(d-2)S^{(0)}_{2},
\en
and for the entropy we get
\be
\mathcal{S}_{2}=S^{(0)}_{2}-\frac{d}{2(d-2)}\mathrm{ln}\, S^{(0)}_{2}+....\label{eq:correcuno}
\en
This result is identical to the one obtained for the AdS-Schwarzschild black holes\cite{das}. 

We can also study the correction to the entropy near the transition point $r_{c}$ (after 
the minimum of $C_{2}$). Assuming $r^{2}_{+}-r^{2}_{c}\approx l^{2}$(small $l$), the entropy behaves as  
\be
S^{(0)}_{2}=\frac{4\pi k_{B}}{G_{2}}r^{d-4}_{+}\left[\frac{1}{(d-4)}+\frac{2(d-3)}{(d-2)(d-1)} \right].
\en
Consequently the relations between the thermodynamic quantities are
\be
T_{H}=\frac{B}{A^{1/(d-4)}}(S^{(0)}_{2})^{1/(d-4)},
\en
\be
C_{2}=\frac{C}{A}S^{(0)}_{2},
\en
where
\be
A=\frac{4\pi k_{B}}{G_{2}}\left[\frac{1}{(d-4)}+\frac{2(d-3)}{(d-2)(d-1)}\right],
\en
\be
B=\frac{1}{8\pi k_{B}}\frac{1}{l^{2}}\left[(d-1)+\frac{(d-5)(d-1)}{2(d-3)}\right],
\en
\be
C=\frac{4\pi k_{B}}{G_{k}}\left[1+\frac{2(d-3)}{d-1}\right]\left[\frac{2(d-3)}{(d-1)}+\frac{(d-5)}{(d-1)}\right].
\en
Therefore
\be
\mathcal{S}_{2}=S^{(0)}_{2}-\frac{d-2}{2(d-4)}\mathrm{ln}\, S^{(0)}_{2}+....
\en
It is interesting to note that the entropy correction in this limit is greater than 
the entropy correction found in (\ref{eq:correcuno}).
\subsection{Chern-Simons Black Holes}
Now let us consider the Chern-Simons black holes with metric (\ref{eq:csbh}). The thermodynamic 
quantities can be obtained considering $d-2k-1=0$. So,
\be
T_{H}=\frac{1}{4\pi k_{B}k}\left((d-1)\frac{r_{+}}{l^{2}}\right),
\en
\be
C_{CS}=k\frac{2\pi k_{B}}{G_{k}}r_{+}\left(1+\frac{r^{2}_{+}}{l^{2}} \right)^{k-1},
\en
and
\be
S_{k}=k\frac{2\pi k_{B}}{G_{k}}\int^{r_{+}}_{0}\left(1+\frac{r^{2}}{l^{2}}\right)^{k-1}dr.
\en
In this case the temperature is not divergent and the specific heat is a continuous monotonically increasing positive function of $r_{+}$

In the limit $r_{+}\gg l$ we get
\be
C_{CS}=(2k-1)S^{0}_{k}
\en
and
\be
\mathcal{S}_{k}=S^{(0)}_{k}-\frac{2k+1}{2(2k-1)}\mathrm{ln}\, S^{(0)}_{k}+.....
\en
Note that in this limit the results, as was expected, are similar to those found in the previous section.

On the other hand, for $r_{+}\approx l$, we obtain
\be
\mathcal{S}_{k}=S^{(0)}_{k}-\frac{3}{2}\mathrm{ln}\, S^{(0)}_{k}+.....
\en
Similar to the proportionality constant found in \cite{das} for the BTZ black hole\cite{banados}. 
\section{Conclusions.}
In this paper we have calculated the entropy corrections for different kinds of 
black holes in the context of higher order gravity. The results are similar 
to those found in the literature despite the fact that the area law is not satisfied. 
Also is confirmed the lack of universality of the logarithmic prefactor. In the case of 
Chern-Simons black holes with small horizon radius we have found that the logarithmic prefactor does not depended of the dimension.

\textbf{ACKNOWLEDGMENT:}

The author thanks Prof. Sandro Silva e Costa at the Physics Department-UFMT for hospitality.
This work was supported
by Funda\c{c}\~ao de Amparo \`a Pesquisa do Estado de
Mato Grosso (FAPEMAT) and Conselho Nacional de Desenvolvimento  
Cient\'{\i}fico e Tecnol\'{o}gico (CNPq). The authors also thank the organizers of the
Conference "$100$ years of Relativity" where these results were presented.


\end{document}